\documentclass{mem}
\usepackage{natbib}\usepackage{txfonts}\usepackage{balance}
\usepackage{graphicx}
\usepackage[a4paper]{hyperref}
\idline{75}{282}
\begin{document}
\def\teff{$T\rm_{eff }$}
\def\kms{$\mathrm {km s}^{-1}$}

\title{
Relativistic orbits and Gravitational
Waves from gravitomagnetic corrections
}

   \subtitle{}

\author{
S. \, Capozziello\inst{1} 
\and M. \, De Laurentis\inst{1}
\and L. \, Forte\inst{1}
\and F. \, Garufi\inst{1}
\and L. \, Milano\inst{1}
          }

  \offprints{S. Capozziello}

\institute{
Dipartimento di Scienze fisiche, Universit\'a
di Napoli `` Federico II'' and INFN Sez. di Napoli, Compl. Univ.
di
Monte S. Angelo, Edificio G, Via Cinthia, I-80126, Napoli, Italy
\email{salvatore.capozziello@na.infn.it}
}

\authorrunning{S. Capozziello }

\titlerunning{Relativistic orbits and GW from gravitomagnetic corrections
}

\abstract{Corrections to the relativistic theory of orbits are discussed 
considering higher order approximations induced by gravitomagnetic
effects. Beside the standard periastron effect of General
Relativity (GR), a new nutation effect was found due to the ${\displaystyle
c^{-3}}$  orbital correction. According to the presence of that new nutation effect we studied  the gravitational waveforms emitted through the capture in a gravitational field of a massive black hole (MBH) of a compact  object (neutron star (NS) or BH) via the quadrupole approximation. We made a numerical study to obtain the emitted gravitational wave (GW) amplitudes. We conclude that the effects we studied could be of interest for the future space laser interferometric GW antenna LISA.

}
\maketitle{}
%%%%%%%%%%%%%%%%%%%%%%%
\section{Introduction}
%%%%%%%%%%%%%%%%%%%%%%%%%
The magnetic field is produced by the motion of electric-charge, i.e. the electric
current. The analogy with gravity consists in the fact that a mass-current can produce
a "gravitomagnetic" field.
The formal analogy between electromagnetic and gravitational fields was
explored by \cite{Einstein}, in the framework of GR, and then by
\cite{Thirring1}. It was shown, by \cite{Thirring}, that a rotating mass generates a
gravitomagnetic field, which in turn, causes the precession of planetary orbits.
We want to study how the relativistic theory of orbits and the production of GW is affected by gravitomagnetic corrections.
The corrections, which are off-diagonal terms in the metric, can be seen as further
powers in the expansion in $c^{-1}$ (up to $c^{-3}$). Nevertheless, the effects on the orbit
behavior involve not only the precession at peri-astron but also nutation corrections.
Our approach suggests that, in the weak field approximation, when considering
higher order corrections in the equations of motion, the gravitomagnetic effects can
be particularly significant. In systems approaching to strong field regimes, these
corrections give rise to chaotic behaviors in the transients dividing stable from
unstable orbits \cite{SMFL}. In general, such contributions are discarded since they are assumed
to be too small but they have to be taken into account as soon as the $\frac{v}{c}$ ratio
is significant.
It is possible to take into account two types of mass-current in gravity. The former
is induced by matter source rotations around the center of mass: it generates the
intrinsic gravitomagnetic field which is closely related to the angular momentum
(spin) of a rotating body. The latter is due to the translational motion of sources.
%%%%%%%%%%%%%%%%%%%%%%%%%%%%%
\section{Gravitomagnetic corrections}
\begin{figure}[ht]
\resizebox{\hsize}{!}{\includegraphics[clip=0.1]{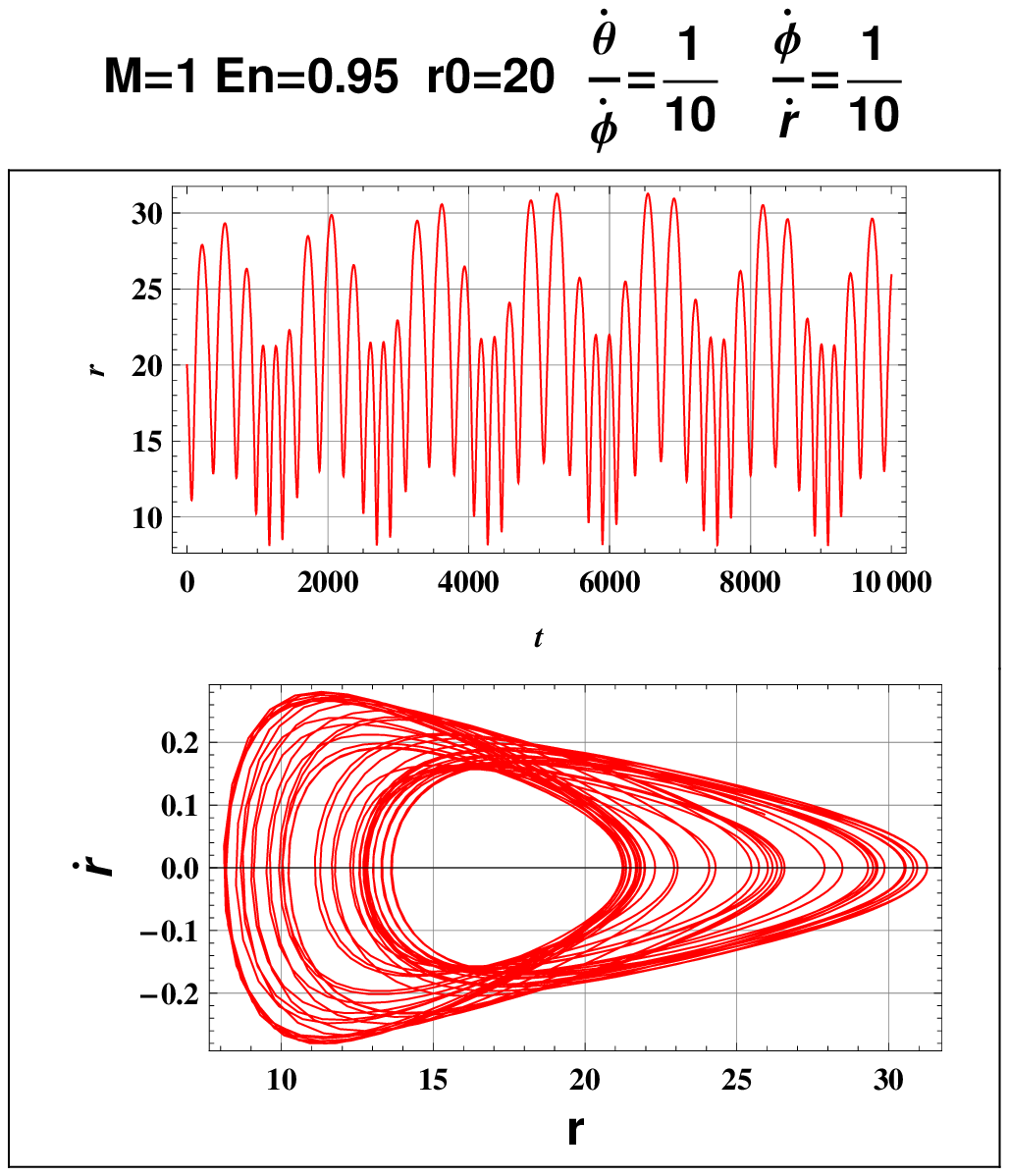}
\includegraphics[clip=0.1]{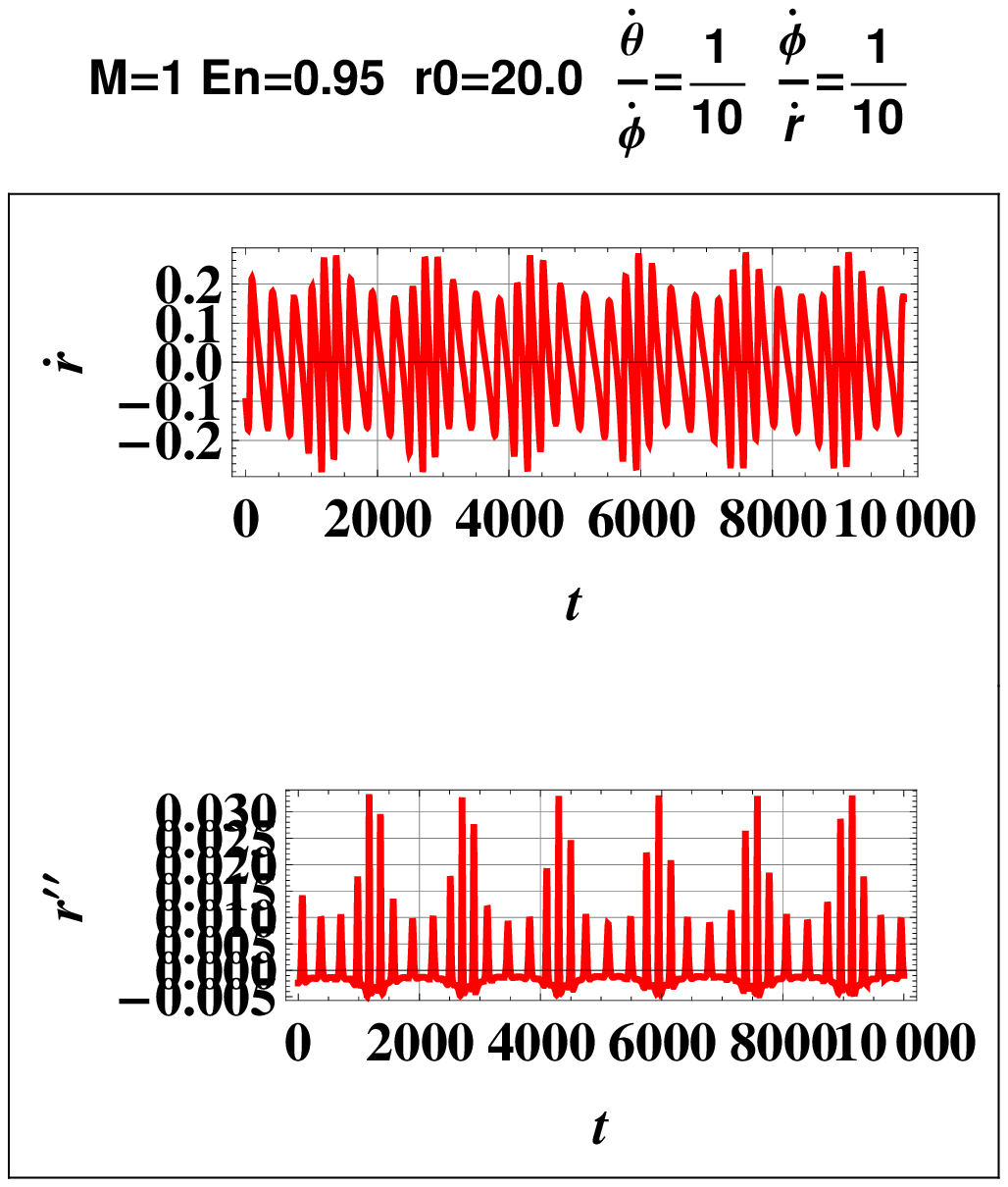}}
\caption {\footnotesize Plots along the panel lines of:  $r(t)$ (upper left), phase portrait of $r(t)$ versus $\dot{r}(t)$ (bottom left), $\dot{r}(t)$ (upper right) and $\ddot{r}(t)$ (bottom right) for a  star  of $1 M_{\odot}$. The examples we are showing were obtained solving the system for the following parameters and initial conditions: $\mu\approx 1 M_{\odot}$, $E=0.95$,$\phi_{0}=0$, $\theta_{0}=\frac{\pi}{2}$, $\dot{\theta_{0}}=\frac{1}{10}\dot{\phi_{0}}$, $\dot{\phi_{0}}=-\frac{1}{10}\dot{r}_{0}$ and $\dot{r}_{0}=-\frac{1}{100}$ and $r_{0}=20 \mu$. The stiffness is evident from the trend of  $\dot{r}(t)$ and $\ddot{r}(t)$}\label{Fig:00}
\end{figure}

%%%%%%%%%%%%%%%%%%%%%%%%%%%%%
Starting from the Einstein field equations in the weak field approximation, one
obtains the gravitomagnetic equations and then the corrections in the metric \cite{gravitation, capozzre}:
\begin{eqnarray}
&& ds^{2}=\left(1+\frac{2\Phi}{c^{2}}\right)c^{2}dt^{2}-\frac{8\delta_{lj}V^{l}}{c^{3}}cdtdx^{j}\nonumber\\ &&-\left(1-\frac{2\Phi}{c^{2}}\right)\delta_{lj}dx^{i}dx^{j}\;.\label{eq:ds_DUE}\end{eqnarray}
By calculating the affine connection related to the metric (\ref{eq:ds_DUE}),
one also obtain the geodesic equations
\begin{equation}
\ddot{x}^{\alpha}+\Gamma_{\mu\nu}^{\alpha}\dot{x}^{\mu}\dot{x}^{\nu}=0\;,\label{eq:geodedica_uno}\end{equation}
where the dot indicate differentiation with respect to the affine parameter. 
 In order to put in evidence the gravitomagnetic
contributions, let us explicitly calculate the Christoffel symbols
at lower orders. By some straightforward calculations, one gets
\begin{equation}\begin{array}{cl}
\Gamma^0_{00} &=0\\
\Gamma^0_{0j} &=\frac{1}{c^2}\frac{\partial\Phi}{\partial x^j} \\
\Gamma^0_{ij} &=-\frac{2}{c^3}\left(\frac{\partial V^i}{\partial x^j}+\frac{\partial V^j}{\partial x^i}\right) \\
\Gamma^k_{00} &= \frac{1}{c^2}\frac{\partial\Phi}{\partial x^k}\\
\Gamma^k_{0j} &=\frac{2}{c^3}\left(\frac{\partial V^k}{\partial x^j}-\frac{\partial V^j}{\partial x^k}\right) \\
\Gamma^k_{ij} &= -\frac{1}{c^2}\left(\frac{\partial \Phi}{\partial
x^j}\delta^k_i+\frac{\partial \Phi}{\partial
x^i}\delta^k_j-\frac{\partial \Phi}{\partial
x^k}\delta_{ij}\right)\end{array}\end{equation} In the
approximation which we are going to consider, we are retaining
terms up to the orders $\Phi/c^2$ and $V^j/c^3$. It is important
to point out that we are discarding terms like
$(\Phi/c^4)\partial\Phi/\partial x^k$,
$(V^j/c^5)\partial\Phi/\partial x^k$, $(\Phi/c^5)\partial
V^k/\partial x^j$, $(V^k/c^6)\partial V^j/\partial x^i$ and of
higher orders. Our aim is to show that, in several  cases like in
tight binary stars, it is not correct to discard higher order
terms in $v/c$ since physically interesting effects could come
out.
The geodesic equations up to $c^{-3}$ corrections are then
\begin{eqnarray}
&& c^{2}\frac{d^{2}t}{d\sigma^{2}}+\frac{2}{c^{2}}\frac{\partial\Phi}{\partial
x^{j}}c\frac{dt}{d\sigma}\frac{dx^{j}}{d\sigma}\nonumber\\ &&
-\frac{2}{c^{3}}\left(\delta_{im}\frac{\partial
V^{m}}{\partial x^{j}}+\delta_{jm}\frac{\partial V^{m}}{\partial
x^{i}}\right)\frac{dx^{i}}{d\sigma}\frac{dx^{j}}{d\sigma}=0\;,\label{time}
\end{eqnarray}
for the time component, and
\begin{eqnarray}
&&\frac{d^{2}x^{k}}{d\sigma^{2}}+\frac{1}{c^{2}}\frac{\partial\Phi}{\partial
x^{j}}\left(c\frac{dt}{d\sigma}\right)^{2}+\nonumber\\ &&
\frac{1}{c^{2}}\frac{\partial\Phi}{\partial x^{k}}\delta_{ij}\frac{dx^{i}}{d\sigma}\frac{dx^{j}}{d\sigma}-\frac{2}{c^{2}}\frac{\partial\Phi}{\partial
x^{l}}\frac{dx^{l}}{d\sigma}\frac{dx^{k}}{d\sigma}+\nonumber \\ && \frac{4}{c^{3}}\left(\frac{\partial
V^{k}}{\partial x^{j}}-\delta_{jm}\frac{\partial V^{m}}{\partial
x^{k}}\right)c\frac{dt}{d\sigma}\frac{dx^{i}}{d\sigma}=0\;,\label{eq:dduexk}\end{eqnarray}
for the spatial components.
Considering only the spatial components, we obtain the orbit equations.
Calling $dl_{euclid}=\delta_{ij}dx_{i}dx_{j}$, and $e^{k}=\frac{dx^{k}}{dl_{euclid}}$ we have, in vector form,
\begin{eqnarray}
&& \frac{d\mathbf{e}}{dl_{euclid}}=-\frac{2}{c^2}\left[\nabla\Phi-\mathbf{e}(\mathbf{e}\cdot\nabla\Phi)\right]+\nonumber\\ &&\frac{4}{c^3}
\left[\mathbf{e}\wedge(\nabla\wedge\mathbf{V})\right]\label{vector}\,.
\end{eqnarray}
The gravitomagnetic term is the second one in Eq.(\ref{vector})
and it is usually discarded since considered not relevant. This is
not true if $v/c$ is quite large as in the cases of tight binary
systems or point masses approaching to BH.
From the above equations we can write the Lagrangian and derive the orbital equations of motions starting from the Euler-Lagrange equations (see \cite{SMFL}.
Our aim is to study how gravitomagnetic
effects modify the orbital shapes and what
are the parameters determining the stability
of the problem.
The energy, the mass and the angular momentum, essentially,
determine the stability.
Beside the standard periastron precession of GR, a nutation
effect is induced by gravitomagnetism and stability depends on it.
The solution of the system of differential equations (ODE) of motion presents some
difficulties since the equations are stiff.
For our purposes, we have found solutions by using the so called
Stiffness Switching Method to provide an automatic mean of switching
between a non-stiff and a stiff solver coupled with a more conventional
explicit Runge-Kutta method for the non-stiff part of differential equations.
Time series of both
$\dot{r}(t)$ and $\ddot{r}(t)$ together with
the phase portrait $r(t), \dot{r}(t)$ are shown
assuming as initial values of the angular
precession and nutation velocities
integer ratios with the radial velocity. In Fig. \ref{Fig:00} we show as example the one with:
$\dot{\varphi}=\frac{1}{10}\dot{r}$ and 
$\dot{\theta}=\frac{1}{10}\dot{\varphi}$
%%%%
%%%%%%%

%%%%%%%%%%%%%%%%%%%%%%%%%%%%%%%%%%%% 
\subsection{Orbits with gravitomagnetic corrections}
\begin{figure}[!ht]
\resizebox{\hsize}{!}{
\includegraphics[clip=18]{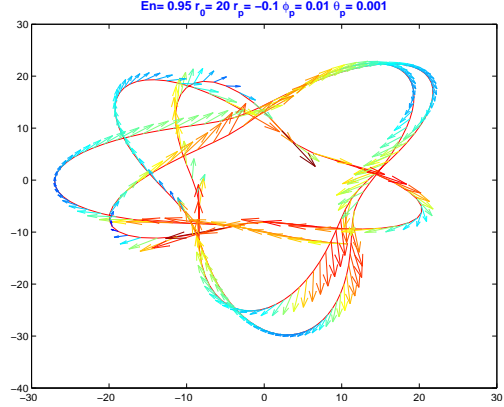}}
\caption{\footnotesize Plots of the basic orbits  with the associated field velocities in false colours}\label{Fig:0a}\end{figure}

\begin{figure}[t!]
\resizebox{\hsize}{!}{\includegraphics[clip=true]{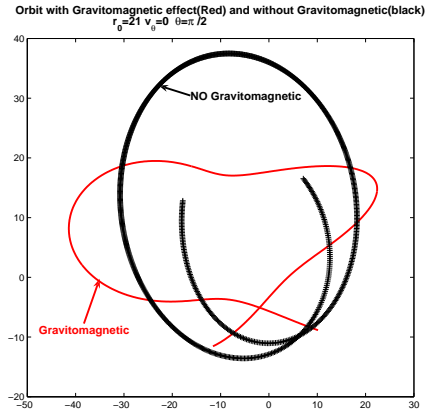}}
\caption { \footnotesize The orbit with gravitomagnetic correction (red-line) and without gravitomagnetic correction (black-line). }\label{Fig:0b}
\end{figure}

\begin{figure}[t!]
\resizebox{\hsize}{!}{\includegraphics[clip=true]{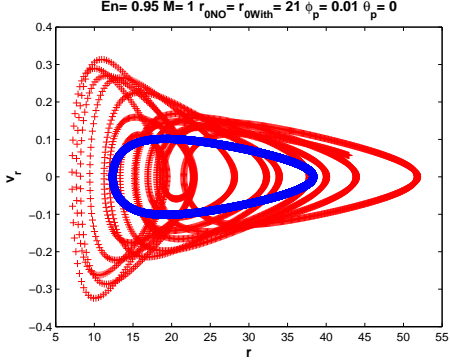}}
\caption { \footnotesize The  phase portrait with (red-line) and without (blue-line) gravitomagnetic orbital correction. The example we are showing was obtained solving the system for the following parameters and initial conditions: $\mu\approx 1 M_{\odot}$, $E=0.95$,$\phi_{0}=0$, $\theta_{0}=\frac{\pi}{2}$, $\dot{\theta_{0}}=\frac{1}{10}\dot{\phi_{0}}$,$\dot{\phi_{0}}=-\frac{1}{10}\dot{r}_{0}$ and $\dot{r}_{0}=-\frac{1}{10}$ and $r_{0}=20 \mu$.}\label{Fig:0c}
\end{figure}

\begin{figure}[t!]
\resizebox{\hsize}{!}{\includegraphics[clip=0.5]{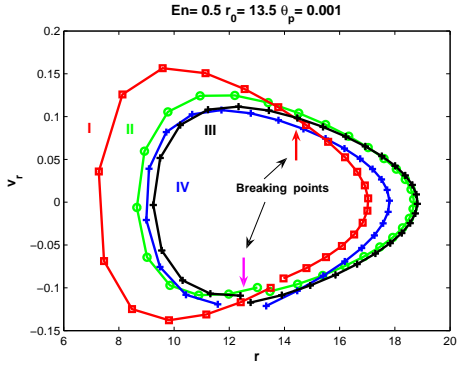}}
\caption {\footnotesize  We show the first four orbits in the phase plane: the red one is labelled I, the green is II , the black is III and the blue is IV. As it is possible to see, the orbits in the phase plane are not closed and they do not overlap the orbital closure point; we called these features 'breaking points'.In this dynamical situation, a small perturbation can lead the system to a transition to chaos as prescribed by the Kolmogorov-Arnold-Moser (KAM) theorem.}\label{Fig:0d}
\end{figure}

%%%%%%%%%%%%%%%%%%%%%%%%%%%%%%%%%%%%%%
In this section  we show some examples of orbits. In Fig.\ref{Fig:0a} are plotted some basic orbits  with the associated field velocities in false colours. Then in Fig. \ref{Fig:0b} we show the orbits with gravitomagnetic correction (red-line) and without gravitomagnetic correction  (black-line). Finally in the Fig. \ref{Fig:0c} there is the phase portrait with (red-line) and without (blue-line) gravitomagnetic orbital correction respectively. The example we are showing was obtained solving the system for the following parameters and initial conditions: $\mu\approx 1 M_{\odot}$, $E=0.95$, $\phi_{0}=0$, $\theta_{0}=\frac{\pi}{2}$, $\dot{\theta_{0}}=\frac{1}{10}\dot{\phi_{0}}$, $\dot{\phi_{0}}=-\frac{1}{10}\dot{r}_{0}$ and $\dot{r}_{0}=-\frac{1}{10}$ and $r_{0}=20 \mu$.
%%%
%%%
%%%%%
%%%%%%%%%%%%%%%%%%%%
In figure \ref{Fig:0d} we show some breaking points of the orbital motion with gravitomagnetic corrections. 
%%%%%
%%%%%
\begin{figure}[t!]
\resizebox{\hsize}{!}{\includegraphics[clip=0.5]{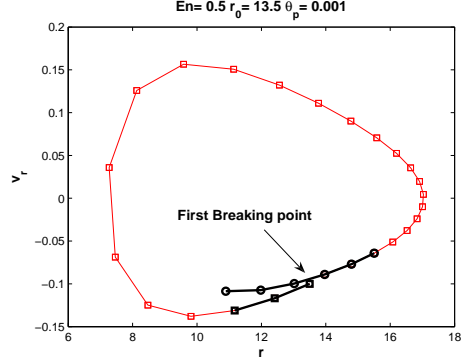}}
\caption {\footnotesize In this figure it is shown the initial orbit with the initial(squares) and final(circles) points marked in black.}\label{Fig:0e}
\end{figure}
%%%%%%%%%%%%%%%%%%%%%%%%%%%%
\section{Gravitational wave in the quadrupole
approximation}
\begin{figure}[t!]
\resizebox{\hsize}{!}
{\includegraphics[clip=true]{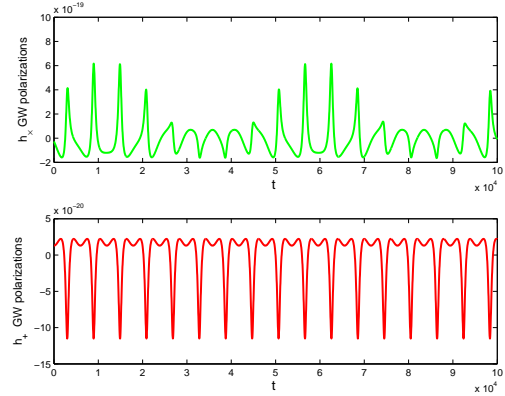}}
\caption{ \footnotesize  The gravitational waveform polarizations $h_{+}$ and $h_{\times}$ for a neutron star (NS) of $1.4 M_{\odot}$}
\label{Fig:0g}\end{figure}

%%%%%%%%%%%%%%%%%%%%%%%%%%%%
Now, considering the orbital equations (see \cite{SMFL}), we know tha direct signatures of gravitational radiation are its amplitude and
its wave-form  \citep{gravitation}. In other words, the identification of a GW signal
is strictly related to the accurate selection of  the shape of
wave-forms by interferometers or any possible detection tool. Such
an achievement could give information on the nature of the GW
source, on the propagating medium, and , in principle, on the
gravitational theory producing such a radiation.
It is well known that the amplitude of GWs can be evaluated by
\begin{equation}
h^{jk}(t,R)=\frac{2G}{Rc^4}\ddot{Q}^{jk}~, \label{ampli1}
\end{equation}
$R$ being the distance between the source and the observer and
$\{j,k\}=1,2$, where $Q_{ij}$ is the  quadrupole mass tensor
\begin{equation}
Q_{ij}=\sum _a
m_a(3x_a^ix_a^j-\delta_{ij}r_a^2)~.\label{qmasstensor}
\end{equation}
Here  $G$ is the Newton constant,  $r_a$  the modulus of the vector
radius of the $a-th$ particle and the sum running over all masses
$m_{a}$ in the system.
At this point we computed the amplitude components with gravitomagnetic corrections in geometrized units (\cite{SMLFL}). 
We performed the numerical simulations in two cases: i) a NS of  $1.4M_\odot$
orbiting around a Super-MBH ( e.g. Sagittarius A*  $10^6M_\odot$) ii) a BH of  $10M_\odot$
orbiting around a Super-MBH.
We considered the reduced mass $\mu=\frac{m_1M_2}{m_1+M_2}$. Computations are performed
with orbital radii measured in mass units.
Initial distances are sampled to show orbits from high
eccentricity up to circularity ($e=\frac{r_{max}-r_{min}}{r_{max}+r_{min}}$).
\begin{table*}
\caption{Data for GW for a NS of  $1.4M_\odot$ orbiting around a Super-MBH}
\label{abun}
\begin{center}
\begin{tabular}{lccccccc}
\hline
\\
 $\frac{r_{0}}{\mu} $ & $  e $ &   $   f(mHz) $  & $ h $ &$  h_{+} $ &  $  h_{\times} $ \\
\hline
\\
  $20  $ & $   0.91 $ &  $   7.7\cdot 10^{-2} $ & $ 2.0\cdot 10^{-22} $ & $  5.1\cdot 10^{-23} $ & $  5.1\cdot 10^{-22} $\\ 
         $200  $ & $  0.79 $ &  $1.1\cdot 10^{-1} $ & $ 1.2\cdot 10^{-20} $ & $  2.2\cdot 10^{-21} $ & $  3.1\cdot 10^{-20} $\\
         $500  $ & $  0.64 $ & $1.4\cdot 10^{-1}$ & $  6.9\cdot 10^{-20}$ & $   8.7\cdot 10^{-21}$ & $   1.7\cdot 10^{-19}$\\ 
        $1000  $ & $ 0.44 $ & $  1.9\cdot 10^{-1} $ & $ 2.6\cdot 10^{-19} $ & $  6.4\cdot 10^{-20}  $ & $ 6.4\cdot 10^{-19} $\\
        $1500  $ & $ 0.28 $ & $  2.3\cdot 10^{-1} $ & $ 4.8\cdot 10^{-19} $ & $  3.6\cdot 10^{-20} $ & $  1.2\cdot 10^{-18} $\\   
        $2000  $ & $ 0.14 $ & $  2.7\cdot 10^{-1} $ & $ 5.9\cdot 10^{-19} $ & $   4.9\cdot 10^{-20} $ & $  1.3\cdot 10^{-18} $\\ 
        $2500   $ & $ 0.01 $ & $   3.1\cdot 10^{-1} $ & $ 5.9\cdot 10^{-19} $ & $  1.7\cdot 10^{-20} $ & $  9.2\cdot 10^{-19} $\\
\hline
\end{tabular}
\end{center}
\end{table*}
In Fig. \ref{Fig:0e}-\ref{Fig:0g} we show respectively the  field velocities of the orbits along the axes of maximum covariances, the total gravitational emission waveform $h$ and  the gravitational waveform polarizations $h_{+}$ and $h_{\times}$ for a NS of $1.4 M_{\odot}$. The waveform were computed for the Earth-distance from Sagittarius A (central Galactic Black Hole). We obtained the numerical examples solving the ODE system  for the following parameters and initial conditions: $\mu\approx1.4 M_{\odot}$, $E=0.95$,$\phi_{0}=0$, $\theta_{0}=\frac{\pi}{2}$,$\dot{\theta_{0}}=0$, $\dot{\phi_{0}}=-\frac{1}{10}\dot{r}_{0}$ and $\dot{r}_{0}=-\frac{1}{100}$ and $r_{0}=(20\div 1000)\mu$. See also Fig. \ref{Fig:0e}-\ref{Fig:0g} and Table 1.
%%%%%%
\begin{figure}[!ht]
\resizebox{\hsize}{!}
{\includegraphics[clip=true]{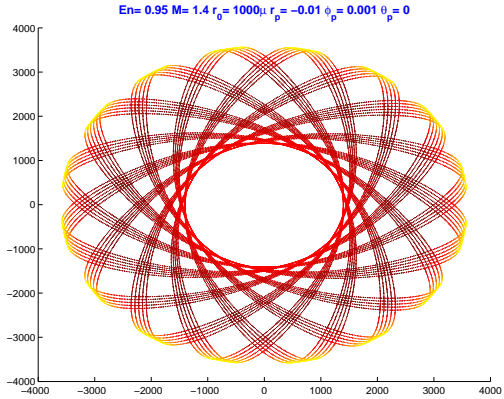}}
\caption { \footnotesize Plot of field velocities of the orbits. }\label{Fig:0e}
\end{figure}
%%%%%
\begin{figure}[t!]
\resizebox{\hsize}{!}
{\includegraphics[clip=true]{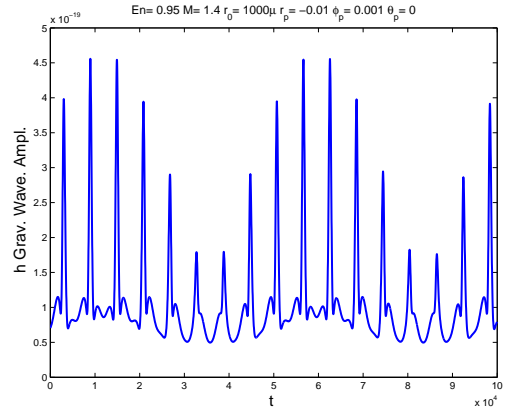}}
\caption{ \footnotesize Total gravitational emission waveform $h$ for a neutron star of  $1.4 M_{\odot}$}
\label{Fig:0f}\end{figure}
%%%%%
%%%%%

\begin{figure}[t!]
\resizebox{\hsize}{!}
{\includegraphics [clip=true]{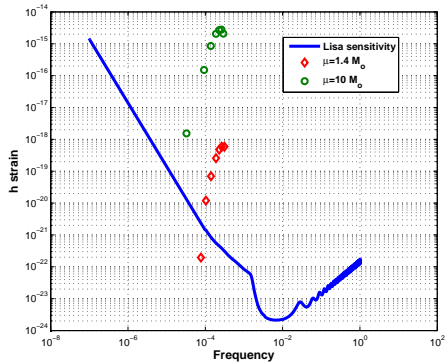}}
\caption { \footnotesize Plot of estimated mean values of gravitational emission in terms of strain $h$ for two binary sources from the galactic center with reduced mass ratio $\mu\approx1.4M_{\odot}$ (red diamonds) and $\mu\approx 10M_{\odot}$ (green circles). The blue line is the foreseen LISA sensitivity curve. The waveforms were computed for the Earth-distance to Sagittarius A (central Galactic Black Hole).}\label{Fig:LISA}
\end{figure}

%%%%%%%%%%%%%%%%%%%%%%%%%%%%%%%%%%%%%%%%%%%%%%%%%%%
Finally we show in Fig. \ref{Fig:LISA} the plot of  the estimated $h$ GW-strain-amplitudes
for the considered binary sources at Galactic
Center distance. The blue-line is the foreseen LISA
sensitivity (one year
integration $+$ white dwarf background
noise). The red diamonds ($1.4M\odot$ ) and
the green circles ( $10M\odot$) are the $h$ values
for the systems we have studied.
%%%%%%%%%%%%%%%%%%%%%%%%%%%%%%%%%%%%%%%%%%%%%%%%%%%
\section{Concluding Remarks}
The gravitomagnetic effect could give rise to interesting phenomena in tight
binding systems such as binaries of evolved objects (neutron stars or black
holes). The effects reveal particularly interesting if v/c is in the range $(10^{-1} \div 10^{-3})c$.
They could be important for objects captured and falling toward extremely
massive black holes such as those at the Galactic Center.
 Gravitomagnetic orbital corrections, after long integration time, induce
precession and nutation effects capable of affecting the stability basin of the
orbits. The global structure of such a basin is extremely sensitive to the initial radial velocities and
angular velocities, the initial energy and masses which can determine possible
transitions to chaotic behavior.
In principle, GW emission could present signatures of gravitomagnetic
corrections after suitable integration times in particular for the on going LISA space laser interferometric GW antenna .

%%%%%%%%%%%%%%%%%%%%%%%%%%%%%%%%%%%%%%%%%%%%%%%%%%

\bibliographystyle{aa}

\end{document}